**Direct Quantification of Molecular Helicity-Dependent Effective Magnetic Fields in Chiral Polypeptide/Ferromagnet Heterostructures**

Theo Balland[1], Buddhika Hondamuni[1,2], Shira Yochelis[3], Meital Ozeri[3], Simon Sochiera[1], Angela Wittmann[1], Yossi Paltiel[3], Dmitry Budker[1,2,4,5], Jonas Fransson[6], Mathias Kläui[1,7,*]

[1] *Institute of Physics, Johannes Gutenberg-Universität Mainz, 55099 Mainz, Germany*
[2] *Helmholtz Institute Mainz, Staudingerweg 18, 55128 Mainz, Germany*
[3] *Institute of Applied Physics, The Hebrew University of Jerusalem, Jerusalem 9190401, Israel*
[4] *GSI Helmholtz-Zentrum für Schwerionenforschung GmbH, 64291 Darmstadt, Germany*
[5] *Department of Physics, University of California, Berkeley, California 94720, USA*
[6] *Department of Physics and Astronomy, Uppsala University, Box 516, 752 37 Uppsala, Sweden*
[7] *Center for Quantum Spintronics, Department of Physics, Norwegian University of Science and Technology, 7491 Trondheim, Norway*

**ABSTRACT**. *Chirality-Induced Spin Selectivity (CISS) refers to structural chirality governing spin transport in chiral systems, with experimental signatures reported in both transport experiments and photoemission. Beyond influencing transport, recent experiments indicate that adsorbed chiral molecules can influence the magnetic state of adjacent ferromagnetic layers, suggesting the existence of chirality-induced effects whose origin and quantitative characterization remain lacking. Here, we investigate chirality-induced long-range magnetic interactions in a hybrid molecular/ferromagnetic multilayer beyond the conventional direct transport coupling. We demonstrate the long-range nature by decoupling the adsorbed chiral molecules from the magnet using an insulating spacer. We show that self-assembled monolayers of chiral polypeptides with a helical structure can act on a ferromagnetic multilayer via an enantiomer-dependent effective magnetic field that may be mediated by chiral phonons through the insulating layer. Using Magneto-Optical Kerr Effect (MOKE) microscopy with on-chip reference regions free of molecules, we directly quantify this field by measuring precisely the reproducible shifts of the local hysteresis loops. The extracted effective field reverses sign for opposite enantiomers and correlates with the enantiomeric excess. These results provide a direct and quantitative link between molecular chirality and magnetic response in hybrid heterostructures, establishing an experimentally accessible benchmark for a CISS-related effect.*

## I. INTRODUCTION.

Chirality-Induced Spin Selectivity (CISS) refers to the emergence of spin-dependent electronic transport in chiral systems, with the response depending on the handedness of the underlying structure. [1–6] The CISS effect has been demonstrated in electrical transport measurements across molecular junctions [7–10] as well as in photoelectron spectroscopy experiments [1,11–13], typically involving steady-state or transient charge flow through the chiral medium. Beyond electrical transport, chiral molecules influencing the magnetic properties of adjacent materials have been reported. [14,15] These effects have mainly been attributed to interfacial exchange interactions, leading to changes in the magnetic anisotropy [14,16,17], altered magnetization curves [18], and changes in the magnetic switching behavior. [15] Consequently, even the stability and dynamics of chiral spin textures such as skyrmions are tuned by the adsorption of chiral molecules. [19] While these observations indicate a molecular chirality-dependent interaction, its origin remains unclear, and quantitative analyses are lacking. Additionally, some experimental configurations involve insulating layers between adsorbed chiral molecules and ferromagnetic layers [14], which suppress the direct electronic coupling that has been central to many explanations of electrical transport measurements. [7,10,19] Therefore, new observables are required to quantify additional CISS effects and to provide theoretical descriptions of CISS with experimental benchmarks to go beyond transport-based framework models.

In this study, we directly quantify the effective magnetic field induced by chiral molecules adsorbed on a multilayer stack where an insulating layer separates the molecules from a ferromagnetic layer with perpendicular magnetic anisotropy. Using Magneto-Optical Kerr Effect (MOKE) microscopy, we probe the magnetization reversal and quantify the associated shift in switching fields, providing a direct measure of the molecular chirality-dependent magnetic interaction.

## II. METHODS

The ferromagnetic multilayer used in this study was engineered to be close to the spin reorientation transition to provide high sensitivity to magnetic changes induced by interfacial magnetic perturbations. [20,21] It exhibits perpendicular magnetic anisotropy (PMA) and supports chiral spin textures, which are sensitive to small variations in the magnetic energy landscape. [22] The multilayer stack as shown schematically in Fig. 1(a) consists of Ta (4 nm)/[$Co_{20}Fe_{60}B_{20}$ (0.9 nm)/Ta (0.08 nm)/MgO (2 nm)/Ta (2 nm)]$^2$/Ta (2 nm)/Au (3 nm), deposited on a Si(100) substrate by magnetron sputtering. The MgO layers induce perpendicular magnetic anisotropy (PMA) through orbital hybridization [23], while Ta interlayers tune the anisotropy and contribute to the interfacial Dzyaloshinskii–Moriya interaction. [24,25] The Au capping layer enables selective molecular adsorption. [26] Importantly, the presence of MgO layers in the stack electrically decouples the magnetic layers from the surface, suppressing direct electronic coupling effects between the adsorbed molecules and the ferromagnet. So our designed configuration provides a platform to investigate chirality-induced effects in the absence of charge transport or direct electronic coupling between the chiral molecules and the ferromagnetic layers.

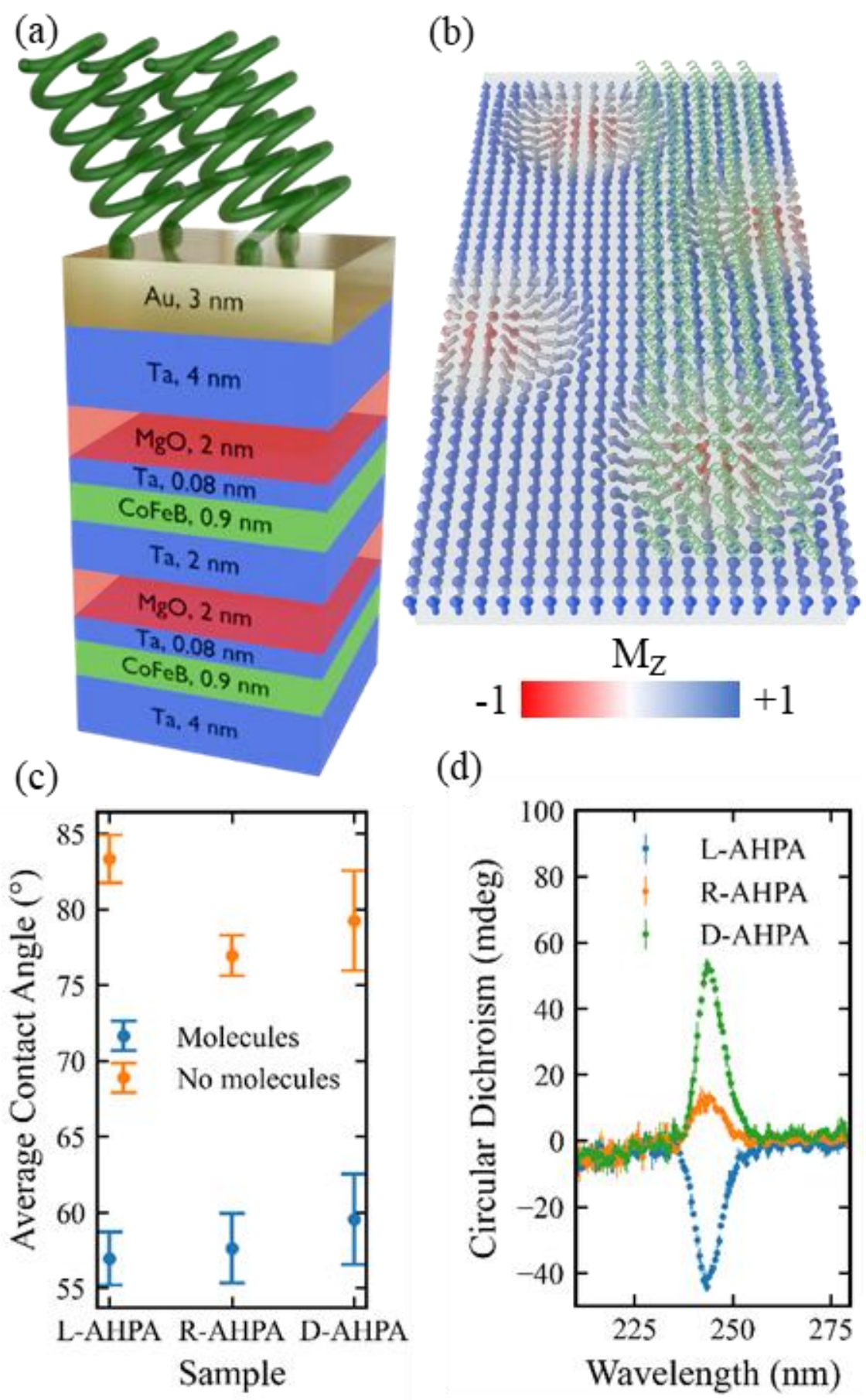

Fig. 1 (a) Schematic of the ferromagnetic multilayer stack with adsorbed chiral molecules. (b) Schematic of skyrmion spin textures under selective adsorption of chiral molecules (green). (c) Contact angle of water droplets deposited on the sample's surfaces with (blue) and without chiral molecule coverage (orange). (d) CD spectra of the prepared solutions of AHPA.

The surface was functionalized with self-assembled monolayers of α-helical polyalanine (AHPA) polypeptides, forming two enantiopure systems (L-AHPA and D-AHPA) and a racemic mixture (R-AHPA). The molecules (molecular mass 3008.6 g/mol, purity 95% and amino-acid sequence: H-[C-(AAAAK)$_7$] -OH possess a terminal cysteine group enabling covalent bonding to the Au surface via thiol surface chemistry. [27] To enable direct comparison under identical experimental conditions, selective adsorption was implemented on each sample as shown in Fig. 1(b). Half of the sample surface was protected using a PMMA layer prior to molecule deposition, resulting in adjacent regions with and without molecular coverage. This approach is critical to eliminate systematic errors arising from sample-to-sample variations, such as differences in magnetic anisotropy, thickness, or temperature conditions. Without such an internal reference, small shifts of the magnetization curve can be masked or misinterpreted due to intrinsic variations between samples. To proceed to the deposition on the chiral polypeptides on the samples' surface, the thin film multilayer stacks were first cleaned using oxygen plasma followed by 10 minutes of immersion in ethanol to remove contaminants and improve surface wettability. Self-assembled monolayers were formed by immersing the samples in 1 mM solutions of AHPA in absolute ethanol, using separate solutions for the L-, D-, and racemic mixtures. All chemical preparation steps were performed in a nitrogen glovebox to avoid contamination. After an overnight self-assembly process, the samples were rinsed with ethanol to remove any unbonded molecules. The PMMA resist was subsequently removed via acetone treatment, yielding samples with well-defined regions with and without molecular coverage. The chirality of the solutions was verified by circular dichroism (CD) spectroscopy prior to the deposition, as shown in Fig. 1(d). [28] The L- and D-AHPA solutions exhibit CD signals of opposite sign, consistent with their opposing handedness, while the racemic mixture exhibits a low CD signal indicating a strong cancellation of the response from the two enantiomers. After the deposition, the molecular coverage was confirmed by comparing the wettability of the bare and molecule-adsorbed regions (Fig. 1(c)). A clear reduction of the water contact angle from ~80° to ~58° is observed after all three adsorptions, consistent with the increased hydrophilicity introduced by the adsorbed molecules.

*Contact author: klaeui@uni-mainz.de

Magneto-optical measurements were performed using a MOKE microscope (evico magnetics) in polar configuration to probe out-of-plane magnetization switching. The magnetic field was applied perpendicular to the sample plane using an electromagnet driven by a Kepco BOP 100-4DL power supply, and the sample temperature was actively stabilized to minimize thermal drift. With a resolution of ~315 nm/px, simultaneous imaging of regions with and without molecular coverage within a single field of view was carried out. Magnetic contrast was obtained by subtracting a background image acquired in a saturated state under a +7 mT field. Magnetization switching was probed by sweeping the out-of-plane magnetic field between −2 mT and +2 mT in steps of 20 µT, with intermediate saturation to ensure a well-defined initial state of each field sweep. At each field step, the magneto-optical signal was imaged and local hysteresis loops were extracted independently from regions with and without molecular coverage. The Kerr signal was quantified by fitting the intensity distribution within each region of interest, yielding the average magnetization signal. The resulting hysteresis loops were corrected for background contributions, normalized, and referenced to the uncovered region to remove residual field offsets. The effective magnetic field was then defined as $H_{eff} = (H_c^+ + H_c^-)/2$, where $H_c^+$ and $H_c^-$ are the positive and negative coercive fields. To ensure robustness, measurements were repeated across multiple regions and multiple cycle repetitions, and the effective magnetic field was extracted systematically. The stability of the external magnetic environment was independently monitored using a nT-sensitivity fluxgate magnetometer to exclude spurious field variations that can lead to artifacts that we can thereby rule out here.

## III. RESULTS

Fig. 2(a) shows the magnetization curve obtained from a representative field of view on the L-AHPA sample, averaged over five repetitions. As chiral molecules can potentially influence several magnetic properties, including magnetic anisotropy and switching behavior, we focus on the coercive fields defined at zero remanent magnetization. A slight variation of magnetic anisotropy is observed between samples, with the D-AHPA sample exhibiting a larger saturation field at 300.0 K. To ensure comparable magnetic conditions, measurements on this sample were performed at a slightly reduced temperature where identical anisotropy values are stabilized (286.6 K). In Fig. 2 (b-d), a zoomed-in view of the magnetization curves reveals a clear shift of the hysteresis loops depending on molecular chirality. The L-AHPA sample exhibits a positive shift of the magnetization curve, while the D-AHPA sample shows a negative shift. In contrast, the racemic sample (R-AHPA) does not display a significant shift within experimental resolution. To ensure the robustness of this observation, this effective magnetic field was extracted systematically from individual measurements across multiple cycle repetitions and different regions of each sample.

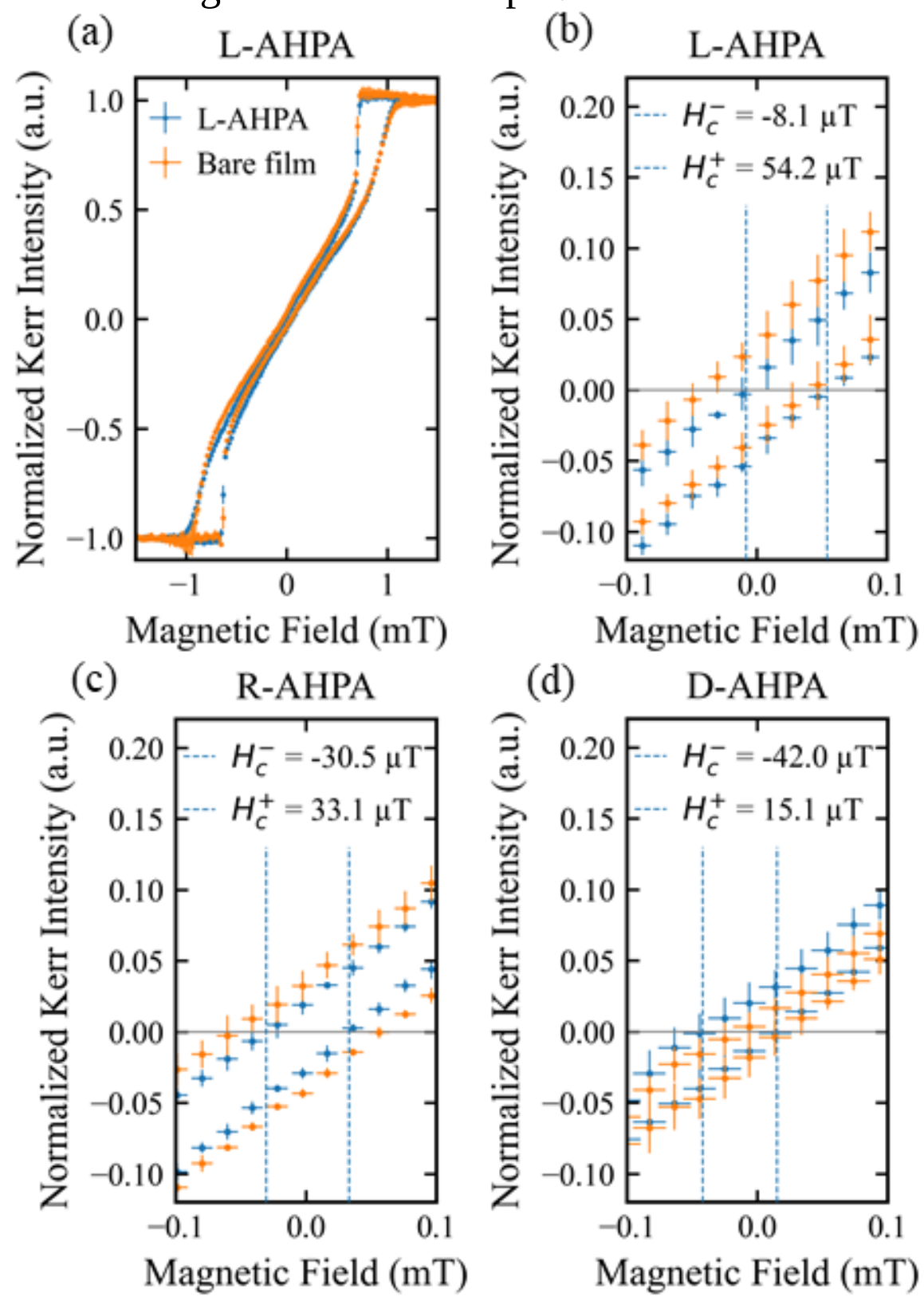

Fig. 2 (a) Magnetic hysteresis curves measured by MOKE signal of the magnetic stack with L-AHPA selective adsorption in a single field of view. (b-d) Zoomed-in view of the MOKE curves for L-AHPA, R-AHPA, and D-AHPA selective adsorption thin films near zero remanent magnetization. The dashed vertical lines are the coercive fields of the thin film with chiral molecule coverage. The MOKE curves of reference magnetic thin film stack (i.e. bare film) are shown in orange for comparison.

The resulting values are summarized in Fig. 3(a). The L-AHPA sample exhibits an average effective magnetic field of +14.4 ± 7.3 µT, while the D-AHPA sample shows an average value of −18.7 ± 7.0 µT. In contrast, the racemic sample yields no significantly changed average field (−5.0 ± 11.3 µT). The standard deviations of the effective magnetic field for the enantiopure samples are comparable, indicating a consistent effect that depends on the molecule's handedness. In contrast, the racemic sample exhibits a broader distribution and fewer data points clustered around the mean value. This is further illustrated in the histogram of extracted effective fields in Fig. 3(c), where the enantiopure samples display well-defined single peaks of opposite sign, whereas the racemic sample exhibits a bimodal distribution with peaks at both positive and negative fields These results demonstrate that chiral molecules induce an effective magnetic field whose sign depends on the molecular handedness, with comparable magnitudes for opposite enantiomers. The reduced average field observed for the racemic sample is consistent with the partial cancellation of contributions of molecules with opposite chirality.

*Contact author: klaeui@uni-mainz.de

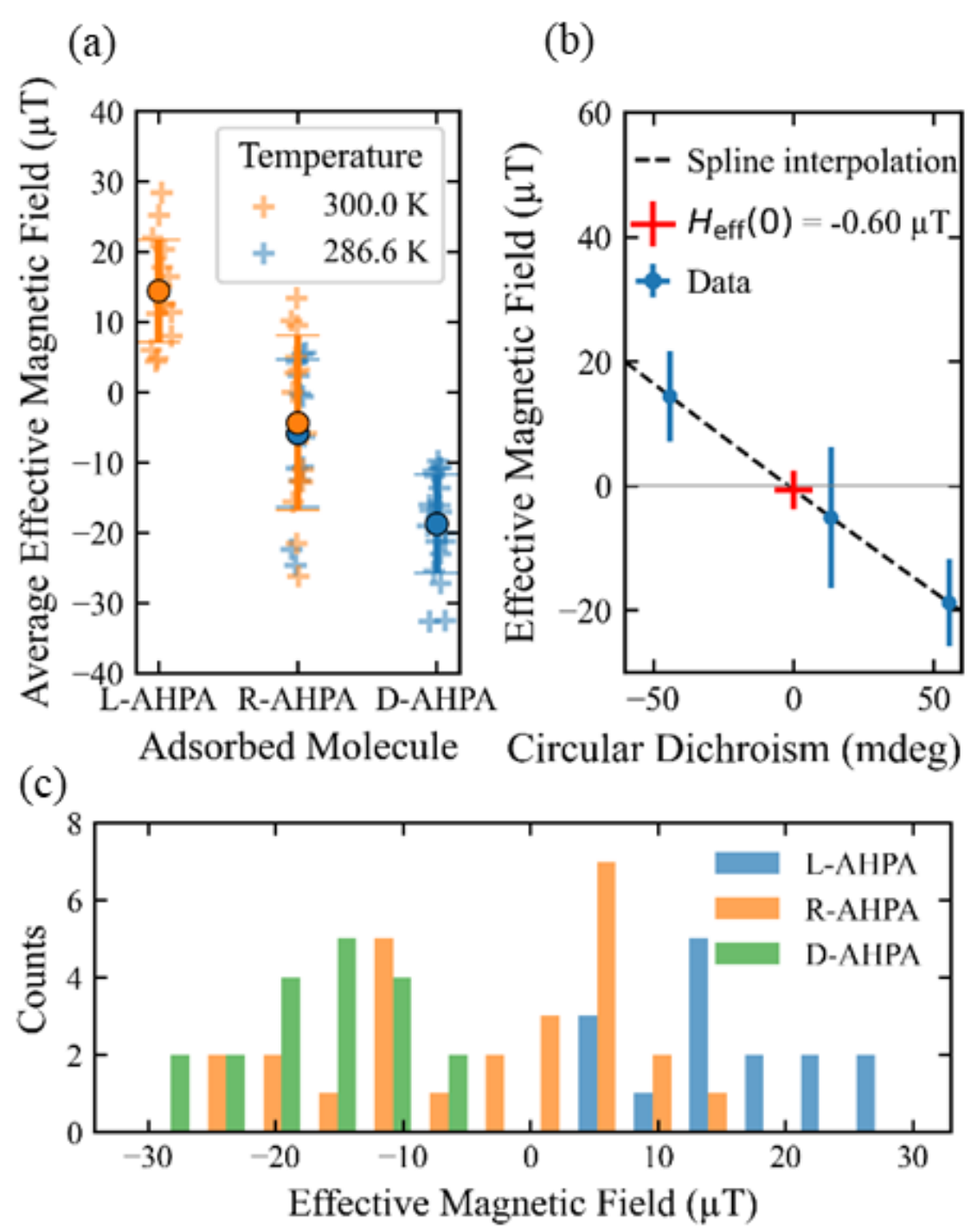

Fig. 3 (a) Average effective magnetic field extracted from single measurements performed on thin films with AHPA selective adsorption at two different temperatures. (b) Correlation between the average effective magnetic field measured across the various thin films and the amplitude of the CD spectra of the AHPA solutions used to deposit chiral molecules onto the surface of the thin films. (c) Distribution of the extracted effective magnetic field from every single magnetization curve measurement.

## IV. DISCUSSION

Comparing the effective magnetic field with the CD spectra measured in solution yields an additional result, shown in Fig. 3(b): the small residual CD signal observed for the racemic mixture indicates a slight enantiomeric excess, which can account for the non-zero average effective field measured on this sample. Upon adsorption, this imbalance leads to a residual net contribution to the interfacial magnetic interaction. The broader and bimodal distribution observed for the racemic sample indicates a more complex interfacial organization compared to the enantiopure monolayers.

Previous studies have reported that racemic chiral molecular mixtures can form distinct self-assembled phases in which opposite enantiomers adsorb in close proximity, leading to different collective electronic or spin-related behavior compared to enantiopure assemblies. [29] Such a mixture-specific response potentially contributes to the peaks at smaller effective magnetic fields observed in the racemic sample. In addition, chirality-dependent adsorption influenced by the local magnetic configuration of the substrate has previously been reported in ferromagnetic systems. [30,31]

Since the present multilayer exhibits a multidomain magnetic state during molecular deposition, local variations of the magnetic texture may influence the adsorption dynamics and lead to spatial fluctuations of the local enantiomeric composition on length scales below the optical resolution. These local variations could influence the magnetization reversal depending on domain nucleation processes and contribute to the statistical broadening and bimodal character of the effective magnetic field distribution. Furthermore, the presence of MgO spacer layers and the approximately 9 nm separation between the adsorbed molecules and the closest ferromagnetic layer prevents direct electronic coupling, ruling out short-range exchange interactions as the mechanism responsible for the observed effective magnetic field.

The responsible interactions have therefore to be long-range to overcome the 9 nm of nonmagnetic layers between the adsorbed chiral polypeptides and the first magnetic layer. The present measurements establish the existence of a molecular chirality-dependent effective magnetic field in the absence of an applied charge current. The sign reversal observed between opposite enantiomers, together with the suppression of the effect in the racemic sample, demonstrates that the measured field is directly linked to molecular handedness rather than to non-enantiomer-specific surface modifications, such as strain. In addition, the architecture of the heterostructure further restricts the range of plausible interaction mechanisms and highlights the need to identify how chirality-dependent effects propagate through electronically decoupled layers.

These observations place important constraints on possible microscopic mechanisms. The results support that chiral molecular adsorption generates a magnetic perturbation acting on a nearby magnetic layer even in the absence of steady-state current flow and without direct electronic coupling.

*Contact author: klaeui@uni-mainz.de

The present observations motivate a phenomenological description in which the adsorbed chiral molecules generate a chirality-dependent effective magnetic, or anisotropy, field acting on the ferromagnetic layer. A possible contribution may arise from a molecular spin density localized at the molecule/metal interface, as predicted by microscopic models of CISS. [32] Such an interfacial spin polarization could in principle generate a local magnetic moment and contribute through dipolar coupling, although the magnitude of this contribution is expected to depend strongly on the separation from the magnetic layer and the spatial distribution of the spin density, as an infinite homogeneous layer of spins does not generate a stray field. Additionally, the interaction may be mediated through spin-lattice coupling. In this scenario, the molecular chirality opens up for anharmonic, or chiral, phonons in the MgO spacer layer. The broken inversion symmetry introduces a non-vanishing average atomic displacement which, in turn, acts on the magnetic moments in the CoFeB layer as an effective magnetic field. [33] Formally, one can write a bilinear coupling between the lattice dynamics and the magnetic moments in the form [33]

$$\mathcal{H}_{MQ} = \int \boldsymbol{M}(\boldsymbol{r}) \cdot \boldsymbol{\mathcal{A}}(\boldsymbol{r}, \boldsymbol{r}') \cdot \boldsymbol{Q}(\boldsymbol{r}') d\boldsymbol{r} d\boldsymbol{r}' \qquad (1)$$

where $\boldsymbol{M}(\boldsymbol{r})$ and $\boldsymbol{Q}(\boldsymbol{r})$ denote the local magnetic moment and nuclear displacement operators, respectively. The coupling $\boldsymbol{\mathcal{A}}(\boldsymbol{r}, \boldsymbol{r}')$ is the electron spin-charge susceptibility which couples to the magnetic moment via local exchange from the Coulomb integral and to the nuclear displacement through electron-phonon coupling.

Phenomenologically, the contribution from Eq. (1) can be regarded as that the nuclear displacement exerts a pseudomagnetic field $g\mu_B \boldsymbol{B}(\boldsymbol{r}) = \int \boldsymbol{\mathcal{A}}(\boldsymbol{r}, \boldsymbol{r}') \cdot \langle \boldsymbol{Q}(\boldsymbol{r}') \rangle$ on the magnetic moment $\boldsymbol{M}(\boldsymbol{r})$. In this way, the structural properties are converted into a field that induces a non-vanishing magnetization in the absence of external magnetic fields. However, unless the average displacement $\langle \boldsymbol{Q}(\boldsymbol{r}) \rangle$ is non-zero, the spin-lattice coupling does not introduce such an offset field. Typically, anharmonic corrections to harmonic oscillations have the capacity to introduce a non-vanishing shift of the average displacement, such that $\langle \boldsymbol{Q}(\boldsymbol{r}) \rangle \neq 0$, which thereby would introduce the magnetic offset field. Since the experimental configuration without chiral molecules has negligible offset, we assume that the overall structure in the absence of chiral molecules, particularly MgO, is well described by harmonic phonons.

Our hypothesis for the acting mechanism is, then, that chiral molecules, by breaking the spatial inversion symmetry, introduce an anisotropy field among the itinerant electrons in the Au/Ta layers which opens up for a spin-dependent electron-phonon coupling to the phonons accommodated in MgO. Through this coupling, an anharmonic correction is forced on the phonons such that the average phononic displacement becomes non-zero. The electron-phonon coupling between itinerant electrons in the Au/Ta layers and the phonons in the MgO is described in detail in the Supplementary. [34]

This mechanism provides a pathway for transmitting molecular chirality across electronically decoupled layers and does not require a finite molecular spin polarization, although both contributions may coexist and reinforce the overall effect. A microscopic formulation of this phenomenological picture could then in a next step provide quantitative constraints on the origin of the measured effective magnetic field.

Finally, we note that our stacks have been shown to host skyrmions [35], which entails an additional advantage: to obtain a localized measure of the field variation beyond the more macroscopic magnetization curve measurements, we propose exploiting the field sensitivity of magnetic skyrmions. The investigated thin film can host magnetic skyrmions, whose equilibrium size is highly sensitive to small magnetic fields. Given the observations, the molecule's handedness is expected to couple to the skyrmion's polarity or core magnetization direction via the induced effective magnetic field. Therefore, at the same absolute applied magnetic field, skyrmions of opposite polarity are expected to have different sizes if the metallic multilayer is covered with chiral polypeptides. This expectation is supported by recent wide-field nitrogen-vacancy magnetometry measurements, which qualitatively observed skyrmion-size variations correlated with both the skyrmion core magnetization direction and the handedness of the adsorbed chiral molecules. [35] So our results open a new path, where measuring the skyrmion size at different magnetic fields can become a highly sensitive technique capable of measuring the molecule-induced local effective magnetic field in the μT range, ignoring effects of domain wall propagation happening during the magnetization reversal.

## V. CONCLUSIONS

In conclusion, we have directly measured an enantiomer-dependent effective magnetic field induced by adsorbed chiral polypeptides in a magnetic multilayer thin film. Using MOKE microscopy with on-chip selective molecular adsorption, we have quantified chiral molecule-induced hysteresis-loop shifts in the μT range and demonstrated opposite signs but comparable field amplitudes for opposite molecular handedness, while racemic mixtures exhibited a strongly attenuated average effect. The measurement of such an effective magnetic field establishes a direct experimental observable linking molecular chirality to a magnetic response in hybrid heterostructures in the absence of applied charge transport. We rule out direct electronic coupling effects by separating the molecules and the magnetic layer by an insulating spacer. One possible mechanism could be related to the coupling between the transient spin generated by the chiral molecule's thermal fluctuations with the ferromagnetic layer through chiral phonons generated in the insulating layer. The information in this case passes through the preservation of angular momentum.

*Contact author: klaeui@uni-mainz.de

The combination of highly sensitive magnetic multilayers and local reference-based microscopy provides a robust experimental framework for quantitatively investigating weak chirality-induced magnetic interactions and constraining microscopic descriptions of CISS-related phenomena.

## ACKNOWLEDGMENTS

This work was supported by the Max Planck Graduate Center with the Johannes Gutenberg University of Mainz (MPGC). The authors acknowledge the funding from Carl Zeiss Stiftung (HYMMS project No. P2022-03-044). Y.P. acknowledges the support from Horizon Europe Marie Sklodowska, grant No. 101071886. The work in Mainz was further supported by the German Research Foundation (DFG, Grant No. 403502522-SPP 2137 Skyrmionics, SFB TRR 173 Spin + X #268565370 (Projects A01, B02 and B14). M.K. acknowledges the support from the European Union's Horizon 2020 research and innovation program under Grant No. 856538 (3D MAGIC), Grant No.101070290 (NIMFEIA), and the TopDyn Center for Dynamics and Topology. J.F. acknowledges Olle Engkvists Stiftelse for financial support. This work was supported in part by the Helmholtz Association (QS4Physics, QT-Batt), and the Deutsche Forschungsgemeinschaft (DFG, German Research Foundation) in the framework to the collaborative research center "Defects and Defect Engineering in Soft Matter" (SFB1552) under Project No. 465145163. This work was supported by the ASPIRE Joint Japan-Germany Quantum Technologies program.

## DATA AVAILABILITY

The data that support the findings of this article are not publicly available. The data are available from the authors upon reasonable request.

[1] K. Ray et al., Science **283**, 814 (1999).
[2] R. Naaman et al., Nat Rev Chem **3**, 250 (2019).
[3] S.-H. Yang et al., Nat Rev Phys **3**, 328 (2021).
[4] D. H. Waldeck et al., APL Materials **9**, 040902 (2021).
[5] J. Fransson, Israel Journal of Chemistry **62**, e202200046 (2022).
[6] B. P. Bloom et al., Chem. Rev. **124**, 1950 (2024).
[7] T. N. H. Nguyen et al., ACS Appl. Mater. Interfaces **14**, 38013 (2022).
[8] Y. Adhikari et al., arXiv:2209.08117.
[9] M. R. Safari et al., Small **20**, 2308233 (2024).
[10] T. N. H. Nguyen et al., Chem. Sci. **15**, 14905 (2024).
[11] B. Göhler et al., Science **331**, 894 (2011).
[12] M. Kettner et al., J. Phys. Chem. C **119**, 14542 (2015).
[13] P. V. Möllers et al., J. Phys. Chem. Lett. **15**, 9620 (2024).
[14] Y. Kapon et al., The Journal of Chemical Physics **159**, 064701 (2023).
[15] O. Ben Dor et al., Nat Commun **8**, 14567 (2017).
[16] N. Sukenik et al., Molecules **25**, 6036 (2020).
[17] R. Sun et al., Science Advances **10**, eadn3240 (2024).
[18] S. Miwa et al., Appl. Phys. Express **13**, 113001 (2020).
[19] Y. Kapon et al., Nano Lett. **25**, 306 (2025).
[20] R. Gruber et al., Nat Commun **13**, 3144 (2022).
[21] R. Gruber et al., Advanced Materials **35**, 2208922 (2023).
[22] M. A. Brems et al., Phys. Rev. Lett. **134**, 046701 (2025).
[23] S. Ikeda et al., Nature Mater **9**, 721 (2010).
[24] J. Cho et al., arXiv:1505.05965 [Cond-Mat] (2015).
[25] K. Everschor-Sitte et al., Journal of Applied Physics **124**, 240901 (2018).
[26] A. Niklewski et al., Langmuir **20**, 8620 (2004).
[27] Y. Xue et al., Nat Commun **5**, 4348 (2014).
[28] S. M. Kelly et al., Biochimica et Biophysica Acta (BBA) - Proteins and Proteomics **1751**, 119 (2005).
[29] T. N. Ha Nguyen et al., ACS Appl. Mater. Interfaces **15**, 17406 (2023).
[30] K. Banerjee-Ghosh et al., Science **360**, 1331 (2018).

*Contact author: klaeui@uni-mainz.de

[31] M. R. Safari et al., Advanced Materials **36**, 2308666 (2024).
[32] J. Fransson, J. Phys. Chem. Lett. **16**, 4346 (2025).
[33] J. Fransson et al., Phys. Rev. Mater. **1**, 074404 (2017).
[34] J. Fransson, Phys. Rev. Res. **5**, L022039 (2023).
[35] B. Hondamuni et al., arXiv:2604.13931.

*Contact author: klaeui@uni-mainz.de

**Direct Quantification of Molecular Helicity-Dependent Effective Magnetic Fields in Chiral Polypeptide/Ferromagnet Heterostructures**

Theo Balland[1], Buddhika Hondamuni[1,2], Shira Yochelis[3], Meital Ozeri[3], Simon Sochiera[1], Angela Wittmann[1], Yossi Paltiel[3], Dmitry Budker[1,2,4,5], Jonas Fransson[6], Mathias Kläui[1,7,*]

[1] *Institute of Physics, Johannes Gutenberg-Universität Mainz, 55099 Mainz, Germany*
[2] *Helmholtz Institute Mainz, Staudingerweg 18, 55128 Mainz, Germany*
[3] *Institute of Applied Physics, The Hebrew University of Jerusalem, Jerusalem 9190401, Israel*
[4] *GSI Helmholtz-Zentrum für Schwerionenforschung GmbH, 64291 Darmstadt, Germany*
[5] *Department of Physics, University of California, Berkeley, California 94720, USA*
[6] *Department of Physics and Astronomy, Uppsala University, Box 516, 752 37 Uppsala, Sweden*
[7] *Center for Quantum Spintronics, Department of Physics, Norwegian University of Science and Technology, 7491 Trondheim, Norway*

**SUPPLEMENTARY**.

The discussion on the anharmonic correction forced on the phonon can be formalized by expanding the displacement operator $\mathbf{Q}(\mathbf{r})$ in plane waves, such that

$$\boldsymbol{Q}(\boldsymbol{r}) = \int l_q \boldsymbol{\varepsilon}_q Q_q e^{i\boldsymbol{q}\cdot\boldsymbol{r}} \frac{dq}{\Omega} \qquad (2)$$

where $\mathrm{q} = (\mu, \mathbf{q})$ denotes the mode number μ and wave vector q, whereas $l_q = \sqrt{\hbar/2\rho\Omega\omega_q}$ is a length scale defined in terms of the phonon energy $\omega_q$, density ρ, and volume Ω. Here, also $Q_q = b_q + b_{\bar{q}}^{\dagger}$ is the quantized form of the displacement, where $b_q$ and $b_{\bar{q}}^{\dagger}$ annihilates and creates a phonon characterized by q [$\bar{q} = (\mu, \bar{\mathbf{q}}) = (\mu, -\mathbf{q})$].

The general form of the electron-phonon coupling is spin-dependent [S1] and can formally be written as

$$\mathcal{H}_{e-ph} = \int \psi^{\dagger}(\boldsymbol{r}) \boldsymbol{\Lambda}_q(\boldsymbol{r}) \psi(\boldsymbol{r}) Q_q \frac{dq}{\Omega} d\boldsymbol{r} \qquad (3)$$

where $\psi(\mathbf{r})$ denotes the spinor for itinerant electrons in the Au/Ta layers. The electron-phonon coupling $\boldsymbol{\Lambda}_q(\mathbf{r}) = \lambda_q(\mathbf{r})\sigma^0 + \boldsymbol{\lambda}_q(\mathbf{r}) \cdot \boldsymbol{\sigma}$ couples phonons to the electron charge via $\lambda_q(\mathbf{r})$ and to the electron spin through $\boldsymbol{\lambda}_q(\mathbf{r})$. In this expansion, $\sigma^0$ and $\boldsymbol{\sigma}$ are the 2 × 2-identity and the vector of Pauli matrices.

We can now analyze the nature of the average displacement and discern which qualities are required for the displacement to be non-vanishing. For the sake of simplicity, we shall make a linear response argument since it is sufficient to reveal the symmetries imposed on the phonons by the environment, in particular by the itinerant electrons in the Au/Ta layers and the chiral molecules.

In linear response theory, assuming stationary conditions, the average displacement is given by

$$\langle Q_q(t) \rangle = -D_{qp}^{r} \int sp \boldsymbol{\Lambda}_{\bar{p}}(\boldsymbol{r}) \boldsymbol{G}^{<}(\boldsymbol{r}, \boldsymbol{r}) \frac{dq}{\Omega} d\boldsymbol{r} \qquad (4)$$

In this expression, $D_{qp}^{r}$ denotes the retarded phonon propagator at zero energy, whereas the lesser Green function $\mathbf{G}^{<}(\mathbf{r}, \mathbf{r})$ is directly proportional to the density of occupied electron states ; *sp* signifies the trace over spin 1/2 space. For the sake of argument, it is sufficient to approximate the phonon propagator with its unperturbed form, $D_{qp}^{r} = \delta_{qp} d_q^{r} = -\delta_{qp} 2/\omega_q$ . Since $\omega_q \neq 0$ for all $\mathbf{q} > 0$, we turn our attention to the electronic properties.

The single-electron Green function $\mathbf{G}(\mathbf{r}, \mathbf{r}') = \langle\langle \psi(\mathbf{r}) | \psi^{\dagger}(\mathbf{r}') \rangle\rangle$ can be expressed as

$$\boldsymbol{G}(\boldsymbol{r}, \boldsymbol{r}') = \boldsymbol{g}(\boldsymbol{r}, \boldsymbol{r}') + \int \boldsymbol{g}(\boldsymbol{r}, \boldsymbol{r}_1) \boldsymbol{\Sigma}(\boldsymbol{r}_1, \boldsymbol{r}_2) \boldsymbol{G}(\boldsymbol{r}_2, \boldsymbol{r}') d\boldsymbol{r}_1 d\boldsymbol{r}_2 \qquad (5)$$

where $\mathbf{g}(\mathbf{r}, \mathbf{r}')$ is the unperturbed Green function and $\boldsymbol{\Sigma}(\mathbf{r}', \mathbf{r}')$ is the self-energy due to interactions and exchange with other particles in the system. As we focus on the influence from the electrons on the phonons, we omit the phonons in the further discussion of the electronic properties. By also omitting the electron-electron interactions among the itinerant electrons, which is justified by strong screening effect, the self-energy comprises the electronic exchange with the chiral molecules. Formally, we assign the expression

$$\mathbf{\Sigma}(\mathbf{r},\mathbf{r}') = \sum_i \mathbf{v}_i\,(\mathbf{r})\mathbf{G}_i\mathbf{v}_i^\dagger(\mathbf{r}'), \tag{6}$$

where the sum runs over all molecules adsorbed on the metal. The hybridization matrix element $\mathbf{v}_\mathrm{i}$ is associated with electron transfer between the metal and the molecule i, whereas $\mathbf{G}_\mathrm{i}$ is the propagator for electrons in the molecule.

Given the Dyson equation, Eq. (5), the lesser Green function can be expressed in terms of the retarded and advanced forms of $\mathbf{G}$ and the lesser self-energy, that is $\mathbf{G}^< = \mathbf{G}^\mathrm{r}\mathbf{\Sigma}^<\mathbf{G}^\mathrm{a}$. This product enables a systematic analysis of the electronic occupation acting on the phonons and, in particular, the properties that lead to phononic anharmonicity.

As a preparation for such analysis, one can notice that any $2\times2$-matrix $\mathbf{A}$ can be expressed on the form $\mathbf{A} = \mathrm{A}_0\sigma^0 + \mathbf{A}_1$, where $\mathrm{A}_0 = sp\mathbf{A}/2$ and $\mathbf{A}_1 = \mathrm{sp}\boldsymbol{\sigma}\mathbf{A}/2$. Physically, $\mathrm{A}_0$ can be associated with the charge properties and $\mathbf{A}_1$ with the spin. These quantities can be further expanded as $\mathrm{A}_0 = \mathrm{A}_{00} + \mathrm{A}_{01}$ and $\mathbf{A}_1 = \mathbf{A}_{10} + \mathbf{A}_{11}$, where the first subscript refers to the (0) charge and (1) spin properties, whereas the latter to spatially (0) symmetric and (1) anti-symmetric properties.

Applying these expansions to the lesser Green function $\mathbf{G}^< = \mathbf{G}^\mathrm{r}\mathbf{\Sigma}^<\mathbf{G}^\mathrm{a}$, see Eq. (A1a), and decomposing into spatially inversion symmetric and anti-symmetric components, we gain access to which anisotropies the chiral molecules contribute to. Especially, the presence of the chiral molecules is legible in the spatially anti-symmetric contribution. It is, furthermore, reasonable to assume that there is no a priori spin-density in the Au/Ta layers, irrespective of whether the surface is decorated with chiral molecules. Then, the spatially inversion-symmetric contribution $\mathrm{A}_{01}(\mathbf{r},\mathrm{q}) = \left[\mathrm{sp}\mathbf{\Lambda}_{\bar{\mathrm{p}}}(\mathbf{r})\mathbf{G}^<(\mathbf{r},\mathbf{r})\right]_{\mathrm{A-S}}$acting on the average displacement is given by the expression

$$\begin{aligned} A_{01}(\boldsymbol{r},q) = 2\int \Sigma_{01}(\boldsymbol{r_1},\boldsymbol{r}_2)\Big\{&\lambda_q^{(00)}(\boldsymbol{r})[G_{00}(\boldsymbol{r},\boldsymbol{r}_1)G_{00}(\boldsymbol{r}_2,\boldsymbol{r}) + \boldsymbol{G}_{11}(\boldsymbol{r},\boldsymbol{r}_1)\cdot\boldsymbol{G}_{11}(\boldsymbol{r}_2,\boldsymbol{r})] \\ &+ \boldsymbol{\lambda}_q^{(11)}(\boldsymbol{r}) \\ &\cdot[\boldsymbol{G}_{11}(\boldsymbol{r},\boldsymbol{r}_1)G_{00}(\boldsymbol{r}_2,\boldsymbol{r}) + G_{00}(\boldsymbol{r},\boldsymbol{r}_1)\boldsymbol{G}_{11}(\boldsymbol{r}_2,\boldsymbol{r}) + i\boldsymbol{G}_{11}(\boldsymbol{r},\boldsymbol{r}_1)\times\boldsymbol{G}_{11}(\boldsymbol{r}_2,\boldsymbol{r})]\Big\} \\ &+ i2\Sigma_{11}(\boldsymbol{r_1},\boldsymbol{r}_2) \\ &\cdot\Big\{\boldsymbol{\lambda}_q^{(11)}(\boldsymbol{r})\times[\boldsymbol{G}_{11}(\boldsymbol{r},\boldsymbol{r}_1)G_{00}(\boldsymbol{r}_2,\boldsymbol{r}) - G_{00}(\boldsymbol{r},\boldsymbol{r}_1)\boldsymbol{G}_{11}(\boldsymbol{r}_2,\boldsymbol{r})]\Big\}d\boldsymbol{r}_1 d\boldsymbol{r}_2 \end{aligned} \tag{7}$$

Here, it is important to appreciate that while $\mathrm{G}_{00}$ accounts for the charge density, $\mathbf{G}_{11}$ indicates the existence of spin currents. These spin currents may be associated with an intrinsic spin-orbit coupling in the metal and/or induced by magnetic defects. In the current context, the spin-orbit coupling of the Au/Ta layers ensures that $\mathbf{G}_{11}$ is non-vanishing, and we can infer that the mechanism in our discussion does not depend on the presence of magnetic defects, although such may modify the characteristics of the effect.

From the expression in Eq. (7), we can read out a few important aspects that help interpreting the experimental results. First, the electronic properties, signified by $\mathrm{G}_{00}$ and $\mathbf{G}_{11}$, indicate that (i) the charge density, $\mathrm{G}_{00}$, and (ii) spin-currents $\mathbf{G}_{11}$, mediate a coupling between the phonons and the chiral molecules. It is also important to notice that the spatially anti-symmetric contribution is non-zero also in the absence of spin-current, which suggests that the magnetic offset discussed here may be present also in set-ups with negligible spin-orbit coupling. Second, the first and second terms in Eq. (7) indicate that the spatially anti-symmetric properties of the chiral molecule, $\Sigma_{01}$, are mediated to the phonons. Lastly, the third term in Eq. (7), which is indicative of spin currents in the molecules, also contributes to the anisotropy. Hence, the spin-orbit coupling associated with chiral molecules is expected to introduce anharmonicity in the phononic character.

*Contact author: klaeui@uni-mainz.de

*Appendix A: $sp\boldsymbol{\Lambda G \Sigma G}$*

$$
\begin{aligned}
sp\boldsymbol{\Lambda G\Sigma G} \quad &= sp(\Lambda_0 G_0 + \boldsymbol{\Lambda}_1 \cdot \boldsymbol{G}_1 + [\Lambda_0 \boldsymbol{G}_1 + \boldsymbol{\Lambda}_1 G_0 + i\boldsymbol{\Lambda}_1 \times \boldsymbol{G}_1] \cdot \boldsymbol{\sigma})(\Sigma_0 G_0 + \boldsymbol{\Sigma}_1 \cdot \boldsymbol{G}_1 \\
&\qquad + [\Sigma_0 \boldsymbol{G}_1 + \boldsymbol{\Sigma}_1 G_0 + i\boldsymbol{\Sigma}_1 \times \boldsymbol{G}_1] \cdot \boldsymbol{\sigma}) \\
&= 2\big((\Lambda_{00} G_{00} + \boldsymbol{\Lambda}_{11} \cdot \boldsymbol{G}_{11})(\Sigma_{00} G_{00} + \boldsymbol{\Sigma}_{11} \cdot \boldsymbol{G}_{11}) + (\Lambda_{00} \boldsymbol{G}_{11} + \boldsymbol{\Lambda}_{11} G_{00} + i\boldsymbol{\Lambda}_{11} \times \boldsymbol{G}_{11}) \\
&\qquad \cdot (\Sigma_0 \boldsymbol{G}_{11} + \boldsymbol{\Sigma}_1 G_{00} + i\boldsymbol{\Sigma}_1 \times \boldsymbol{G}_{11})\big)
\end{aligned} \tag{A1a}
$$

$$
\begin{aligned}
&= 2\Lambda_{00}\big(G_{00}\Sigma_0 G_{00} + G_{00}\boldsymbol{\Sigma}_1 \cdot \boldsymbol{G}_{11} + \Sigma_0 \boldsymbol{G}_{11} \cdot \boldsymbol{G}_{11} + \boldsymbol{G}_{11} \cdot \boldsymbol{\Sigma}_1 G_{00} + i\boldsymbol{G}_{11} \cdot (\boldsymbol{\Sigma}_1 \times \boldsymbol{G}_{11})\big) \\
&\qquad + 2\boldsymbol{\Lambda}_{11} \\
&\qquad \cdot \big(\boldsymbol{G}_{11}\Sigma_0 G_{00} + \boldsymbol{G}_{11}\boldsymbol{\Sigma}_1 \cdot \boldsymbol{G}_{11} + G_{00}\Sigma_0 \boldsymbol{G}_{11} + G_{00}\boldsymbol{\Sigma}_1 G_{00} + iG_{00} \\
&\qquad \cdot (\boldsymbol{\Sigma}_1 \times \boldsymbol{G}_{11})\big) + i2(\boldsymbol{\Lambda}_{11} \times \boldsymbol{G}_{11}) \cdot (\Sigma_0 \boldsymbol{G}_{11} + \boldsymbol{\Sigma}_1 G_{00} + i\boldsymbol{\Sigma}_1 \times \boldsymbol{G}_{11})
\end{aligned}
$$

*Inversion even:*

$$
\begin{aligned}
&= 2\Lambda_{00}\Sigma_{00}(G_{00}G_{00} + \boldsymbol{G}_{11} \cdot \boldsymbol{G}_{11}) + 2\boldsymbol{\Lambda}_{11} \cdot \boldsymbol{\Sigma}_{11}(G_{00}G_{00} - \boldsymbol{G}_{11} \cdot \boldsymbol{G}_{11}) + 2\Lambda_{00}\boldsymbol{\Sigma}_{11} \\
&\qquad \cdot (G_{00}\boldsymbol{G}_{11} + \boldsymbol{G}_{11}G_{00}) + 2\Sigma_{00}\boldsymbol{\Lambda}_{11} \cdot (\boldsymbol{G}_{11}G_{00} + G_{00}\boldsymbol{G}_{11}) \\
&\qquad + 2(\boldsymbol{\Lambda}_{11} \cdot \boldsymbol{G}_{11})(\boldsymbol{\Sigma}_{11} \cdot \boldsymbol{G}_{11}) + 2(\boldsymbol{\Sigma}_{11} \cdot \boldsymbol{G}_{11})(\boldsymbol{\Lambda}_{11} \cdot \boldsymbol{G}_{11}) \\
&\qquad + i2(\boldsymbol{\Lambda}_{11}\Sigma_{00} - \Lambda_{00}\boldsymbol{\Sigma}_{11}) \cdot (\boldsymbol{G}_{11} \times \boldsymbol{G}_{11}) + i2\boldsymbol{\Sigma}_{10} \\
&\qquad \cdot \big(\boldsymbol{\Lambda}_{11} \times (\boldsymbol{G}_{11}G_{00} + G_{00}\boldsymbol{G}_{11})\big)
\end{aligned} \tag{A1b}
$$

*Inversion odd:*

$$
\begin{aligned}
&= 2\Lambda_{00}\Sigma_{01}(G_{00}G_{00} + \boldsymbol{G}_{11} \cdot \boldsymbol{G}_{11}) + 2\boldsymbol{\Lambda}_{11} \cdot \boldsymbol{\Sigma}_{10}(G_{00}G_{00} - \boldsymbol{G}_{11} \cdot \boldsymbol{G}_{11}) + 2\Lambda_{00}\boldsymbol{\Sigma}_{10} \\
&\qquad \cdot (G_{00}\boldsymbol{G}_{11} + \boldsymbol{G}_{11}G_{00}) + 2\Sigma_{01}\boldsymbol{\Lambda}_{11} \cdot (\boldsymbol{G}_{11}G_{00} + G_{00}\boldsymbol{G}_{11}) \\
&\qquad + 2(\boldsymbol{\Lambda}_{11} \cdot \boldsymbol{G}_{11})(\boldsymbol{\Sigma}_{10} \cdot \boldsymbol{G}_{11}) + 2(\boldsymbol{\Sigma}_{10} \cdot \boldsymbol{G}_{11})(\boldsymbol{\Lambda}_{11} \cdot \boldsymbol{G}_{11}) \\
&\qquad + i2(\boldsymbol{\Lambda}_{11}\Sigma_{01} - \Lambda_{00}\boldsymbol{\Sigma}_{10}) \cdot (\boldsymbol{G}_{11} \times \boldsymbol{G}_{11}) + i2\boldsymbol{\Sigma}_{11} \\
&\qquad \cdot \big(\boldsymbol{\Lambda}_{11} \times (\boldsymbol{G}_{11}G_{00} + G_{00}\boldsymbol{G}_{11})\big)
\end{aligned} \tag{A1c}
$$

*Contact author: klaeui@uni-mainz.de

[S1] J. Fransson, Phys. Rev. Res. **5**, L022039 (2023).

*Contact author: klaeui@uni-mainz.de